\documentclass[3p,times]{elsarticle}

\usepackage{ecrc}

\volume{00}

\firstpage{1}

\journalname{Nuclear Physics A}

\runauth{R. Chatterjee}

\jid{nupha}

\usepackage{epsfig} % do not change
\usepackage{amssymb}
\usepackage[figuresright]{rotating}

\begin{document}

\begin{frontmatter}



\dochead{}

\title{Centrality and initial formation time dependence of the emission of thermal photons from fluctuating initial conditions at RHIC and LHC}


\author{Rupa Chatterjee$^{1}$, Hannu Holopainen$^{1,2,3}$, Thorsten Renk$^{1,2}$, and Kari J. Eskola$^{1,2}$}
\address{$^1$Department of Physics, P.O.Box 35, FI-40014 University of Jyv\"askyl\"a, Finland}
\address{$^2$Helsinki Institute of Physics, P.O.Box 64, FI-00014 University of Helsinki, Finland}
\address{$^3$Frankfurt Institute for Advanced Studies, Ruth-Moufang-Str. 1, D-60438 Frankfurt am Main, Germany}

\begin{abstract}
Event-by-event fluctuating initial conditions (IC) in the ideal hydrodynamic calculation are known to enhance the production of thermal photons significantly compared to a smooth initial state averaged profile in the range $p_T >$ 1 GeV/$c$  for 200A GeV Au+Au collisions at RHIC and 2.76A TeV Pb+Pb collisions at LHC. The 'hotspots' or the over-dense regions in the fluctuating IC produce more high $p_T$ photons compared to the smooth IC due to the strong temperature dependent emission of the thermal radiation. This enhancement is expected to be more pronounced for peripheral collisions, for lower beam energies, and for larger values of plasma formation time. A suitably normalized ratio of central to peripheral yield of thermal photons ($R_{cp}^\gamma$) is a potential probe to study the density fluctuations and their size in the initial conditions.
\end{abstract}
%\begin{keyword}
%Event-by-event hydrodynamics, fluctuations, initial conditions, thermal photons.
%\end{keyword}
\end{frontmatter}

Hydrodynamics with event-by-event (e-by-e) fluctuating IC is more realistic than hydrodynamics with smooth IC to model the evolution of the hot and dense matter produced in relativistic heavy ion collisions~\cite{hannu}. Probes which are sensitive to the initial state are especially suitable to study the pattern of fluctuations in the initial QCD matter density distribution. Thermal photons serve this purpose quite well as high $p_T$ ($>$ 1 GeV/$c$) photons are mostly emitted from the initial state (within a few fm) of the system expansion.

We use e-by-e hydrodynamics with fluctuating IC developed in~\cite{hannu} along with state of the art photon rates (plasma rate from~\cite{amy} and hadronic rates from~\cite{trg}) to study the effect of initial state fluctuations on the production of thermal photons from 200A GeV Au+Au collisions at RHIC and 2.76A TeV Pb+Pb collisions at LHC~\cite{chre1,chre2} at different collision centralities. The initial formation time $\tau_0$ is taken as 0.17 fm/$c$ and 0.14 fm/$c$ at RHIC and LHC respectively with a default value of fluctuation size parameter ($\sigma$) 0.4 fm (see ~\cite{chre1} 
and~\cite{chre2} for detail).

Initial state fluctuations are found to enhance the production of thermal photons significantly compared to a smooth initial state averaged profile in the range $2 \le p_T \le 4 $ GeV/$c$ (which is believed to be dominated by thermal radiation in the direct photon spectrum) at RHIC and LHC energies~\cite{chre1, chre2}. The 'hotspots' or the over-dense regions in the fluctuating IC produce more high $p_T$ photons compared to the smooth IC due to the strong temperature dependent emission of the thermal radiation where the emission rates are exponential in temperature. This is an early time effect when the radial flow is still very small. The initial density distribution becomes more homogeneous for larger values of size parameter and we see less enhancement in the production for larger values of $\sigma$~\cite{chre1}.

\begin{figure}
\centering
\includegraphics[height=4.2cm, clip=true]{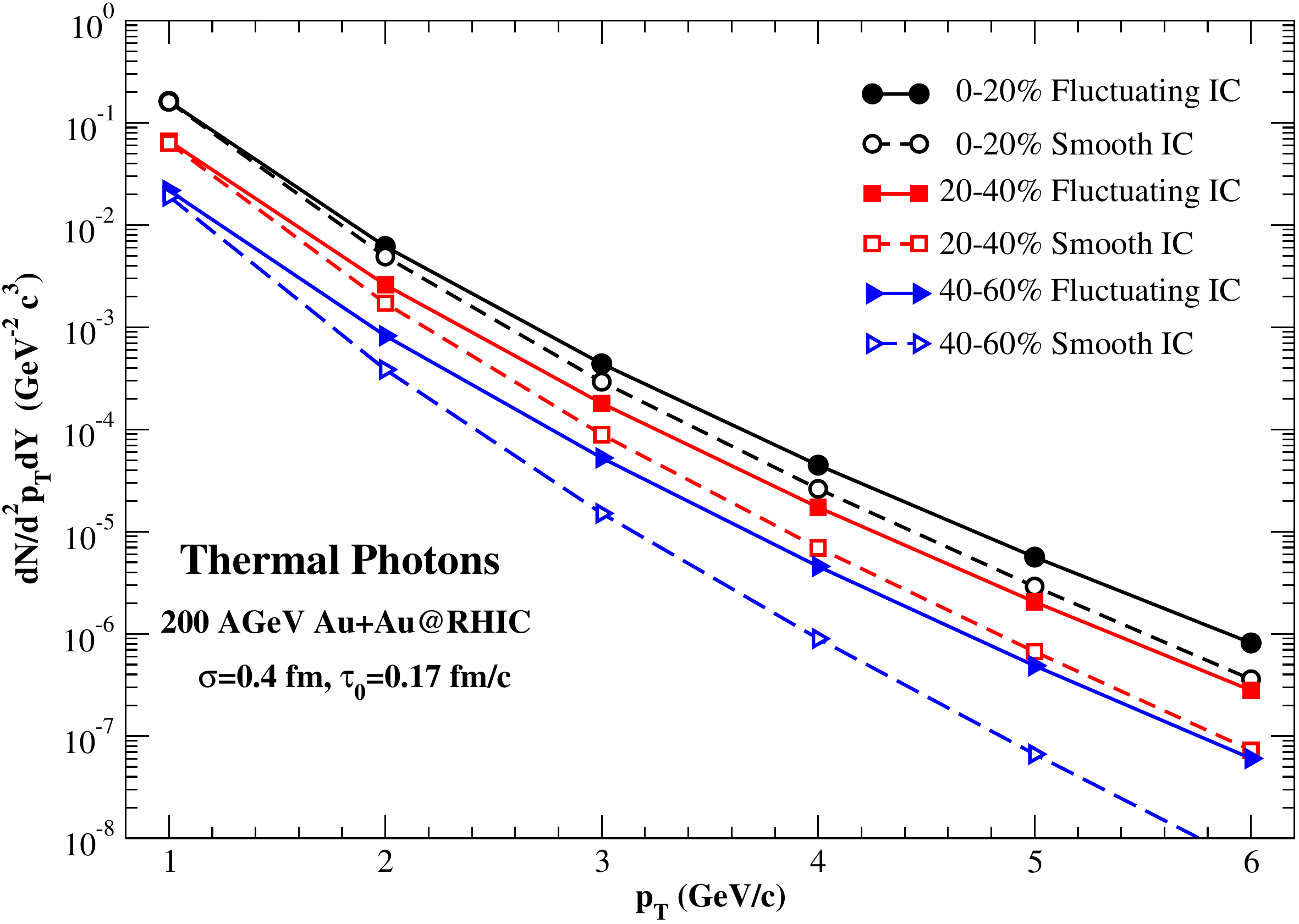}
\includegraphics[height=4.2cm, clip=true]{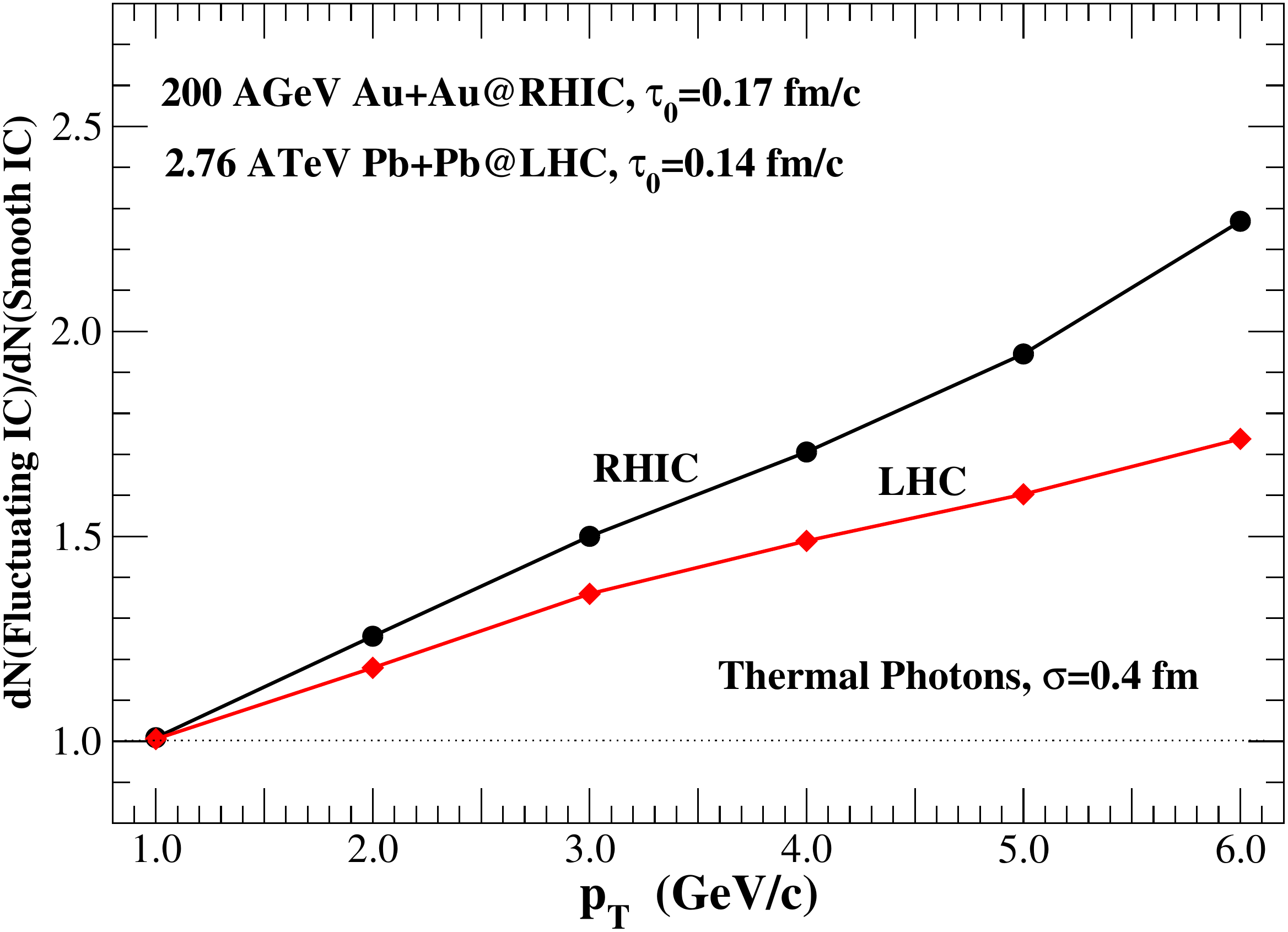}
\includegraphics[height=4.2cm, clip=true]{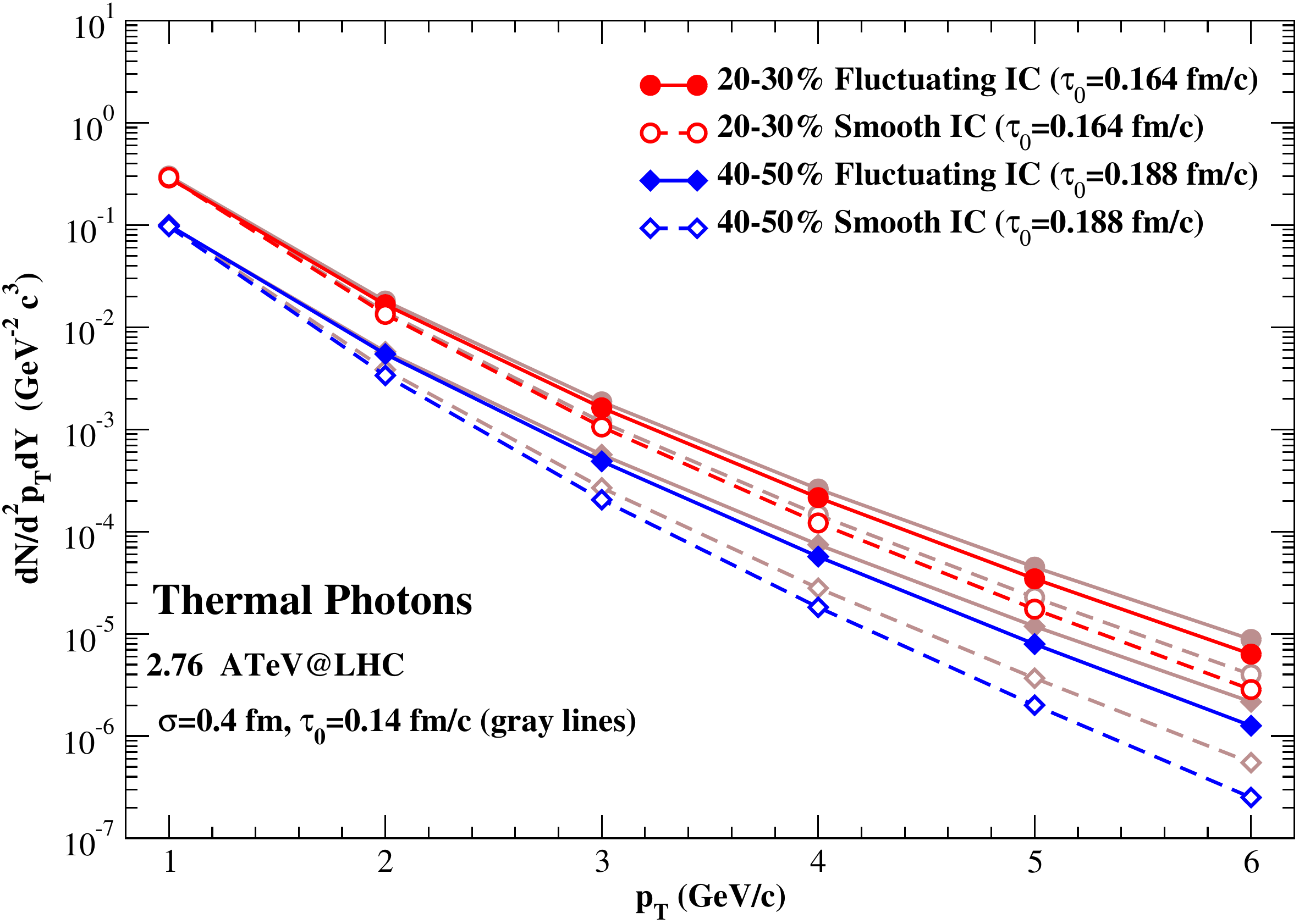}
\includegraphics[height=4.2cm, clip=true]{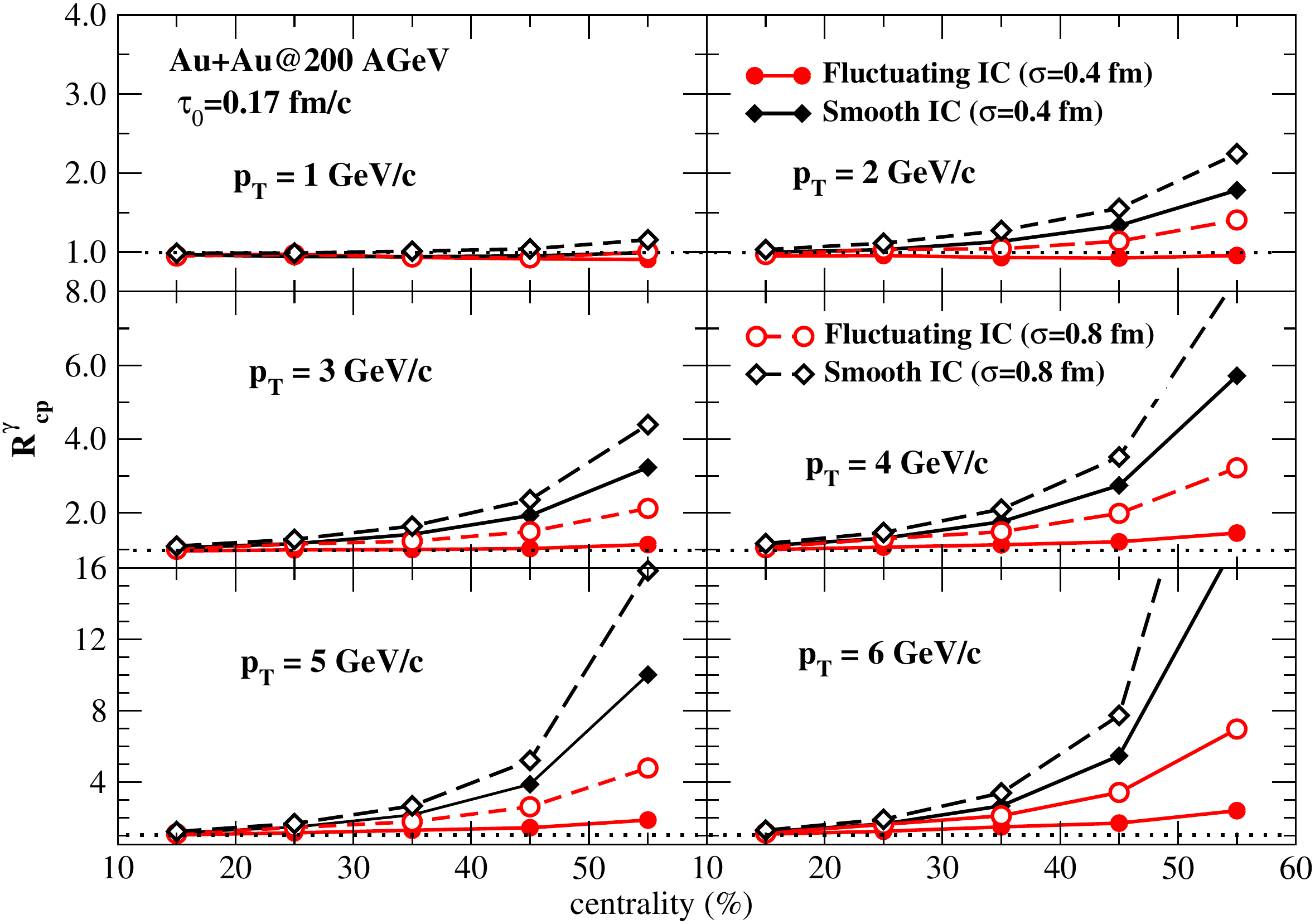}
\caption{[Upper left] $p_T$ spectra of thermal photons from smooth and fluctuating IC at RHIC for different collision centralities. [Upper right] Enhancement due to fluctuations in the IC is more at RHIC than at LHC for same centrality bins. [Lower left] $p_T$ spectra are sensitive to the value of $\tau_0$ which may also vary with collision centralities. [Lower right] Ratio of central to peripheral yield of thermal photons as a function of collision centrality.}
\label{fig1}       % Give a unique label
\end{figure}

We expect that the relative  importance of the 'hotspots' should increase for peripheral collision as the relative importance of the fluctuations increases in peripheral collisions. In detailed studies, the effect of fluctuations is indeed found to be more pronounced for peripheral collision than for central collisions (upper left panel of Figure~\ref{fig1}). The enhancement is less at LHC than at RHIC for the same centrality bin (upper right panel of Figure~\ref{fig1}).

The $p_T$ spectra of thermal photons at RHIC and LHC energies depend strongly on the initial formation time of the plasma which is not known unambiguously and which may also vary with collision centralities. Results from centrality dependent $\tau_0$ values at LHC are shown in lower left panel of Figure~\ref{fig1}. From results at different formation times as well as at centrality dependent $\tau_0$ values  we see that it is difficult to distinguish between the effects of the IC fluctuations and formation time only by looking at the $p_T$ spectra of thermal photons~\cite{chre2}. Thus, one needs to look for some other experimentally measurable quantity where the uncertainties regarding the initial conditions can be reduced. The ratio of central to peripheral yield of thermal photons ($R_{cp}^{\gamma}$) normalized by the number of binary collisions as a function of collision centrality and $p_T$ can be a useful measure of the fluctuation size parameter as shown in lower right panel of Figure~\ref{fig1}.

\vspace{0.2cm}
%\section*{Acknowledgments}
We gratefully acknowledge the financial support by the Academy of Finland. TR and RC are supported by the Academy researcher program (project  130472) and KJE by a research grant (project 133005). HH was supported by the national Graduate School of Particle and Nuclear Physics and the Extreme Matter Institute (EMMI). We acknowledge CSC -- IT Center for Science in Espoo, Finland, for the allocation of computational resources.

\bibliographystyle{elsarticle-num}

\end{document}